%% file: mpp.tex
\input ./jnl.tex

\def\caption#1{\noindent #1}
\ignoreuncited
\singlespace
\title{FACTORING IN A DISSIPATIVE QUANTUM COMPUTER}
\bigskip
\author{Cesar Miquel$^1$, Juan Pablo Paz$^1$ and Roberto Perazzo$^{1.2}$}
\affil{$(1)$: Departamento de F\'\i sica, Facultad de Ciencias Exactas y 
Naturales, Pabell\'on 1, Ciudad Universitaria, 1428 Buenos Aires, 
Argentina}
\affil{$(2)$: Centro de Estudios Avanzados, Universidad de Buenos Aires, 
J.E. Uriburu 850, 1424 Buenos Aires, Argentina}

\abstract{We describe an array of quantum gates implementing 
Shor's algorithm for prime factorization in a quantum computer.  
The array includes a circuit for modular exponentiation with several 
subcomponents (such as controlled multipliers, adders, 
etc) which are described in terms of elementary Toffoli gates. 
We present a simple analysis of the impact of losses and 
decoherence on the performance of this quantum factoring circuit. 
For that purpose, we simulate a quantum computer which is running 
the program to factor $N=15$ while interacting with a dissipative 
environment. As a consequence of this interaction 
randomly selected qubits may spontaneously decay. 
Using the results of our numerical simulations 
we analyze the efficiency of some simple error correction techniques.}
\oneandahalfspace

\subhead{1. Introduction}

In recent years there has been an explosion of activity in the area 
of quantum computation (see \Ref{Seth95,BennettNat}). In part, 
this was a consequence of a very important discovery made in 1994 
by Peter Shor, who 
demonstrated that two problems which are thought to be classically 
intractable (finding prime factors and discrete logarithms of integer 
numbers) could be efficiently solved  
in a quantum computer \refto{Shor1, Shor2}. 
Shor's results added a practical motivation for the study of quantum 
computation which, until that time had received the attention of 
a smaller community of people interested in fundamental aspects of 
quantum mechanics, the physics of information, algorithmic complexity theory, 
etc. By now, quantum computation is a growing field which is developing
not only due to the work of theorists but, fortunately, also due to 
recent advances in experimental techniques. In fact, in the last two years
there have been a few interesting experiments aiming at constructing 
quantum gate prototypes (see \Ref{CiracZoller95,Weinfurter95,Kimble95}).

There are many open questions concerning the mathematics and also the 
physics of quantum computers. In fact, we still don't know what is the 
real power of quantum computation from the algorithmic complexity point of 
view. (Until now, attempts towards demonstrating their usefulness to solve 
NP--complete problems were not successful.) On the other hand, the physics 
of quantum computers also presents many important challenges. 
Among the most important open questions is the  
understanding of the impact of the process of 
decoherence (an issue that attracted some attention over the last two years
\refto{Unruh94,Paz94,CLSZ95,BarencoDeco,ChuangLaf95}). 
Decoherence \refto{ZurekPT} is a physical
process by which the quantum interference effects,  essential 
for the proper functioning of the quantum computer, are destroyed by 
the interaction between the computer and its environment. 
This interaction creates irreducible computer--environment 
correlations which, as the environment is 
unobserved, induce the dynamical collapse of the computer's wave 
function. Decoherence may be potentially devastating but,
as recent studies suggest, 
there may be ways in which one can reduce  
the problem. For that purpose, a few ideas have been advanced. Shor proposed
a procedure for recovering quantum coherence by using coding
\refto{Shor95} (see also \Ref{ChuangLaf95}), 
and similar methods have been proposed for ``purifying'' entangled pairs
before using them for transmiting quantum information through noisy channels 
\refto{Bennettetal95}. This, combined with the possibility of building error 
correction 
schemes based on the ``watchdog'' effect\refto{watchdog} are promising 
ideas that are currently under investigation. 

However, to give a specific answer to 
the question of how important is decoherence for factoring one needs to 
be rather specific. The answer will depend upon the computer implementation
(hardware) and also on the particular algorithm (software) used. 
For example, the possibility of implementing error correction schemes based
on watchdog effect depends upon having a computer evolving 
in such a way that at some known instants it is in a known state (or at least
some qubits are in a known state, so that we can measure them 
without disturbing the computer). 
The aim of this paper is to begin a study on the impact of 
dissipation and decoherence on a quantum factoring computer. For this 
purpose we design a quantum factoring circuit analyzing how 
its performance is affected when the interaction with an environment is
included. 

Several recent papers are related to ours:
Chuang et al. \refto{CLSZ95} 
described on general grounds the potentially devastating  effects that decoherence may have upon a factoring computer. 
Their results, which were obtained using a simple description of the quantum 
computer, which makes no reference to a specific quantum circuit, suggest 
that by having a low enough decay rate and using appropriate 
error correction techniques one may be able to implement factoring in a quantum 
computer. Cirac and Zoller \refto{CiracZoller95} presented a numerical study of 
the effects of errors on the quantum Fourier Transform (FT) 
subroutine, which plays a central role on the factoring program. Their 
simulation was done by considering the effect of spontaneous decay 
while a computer made of cold trapped ions runs the FT program (previously 
designed by Coppersmith and others \refto{Shor2,Copper94}). 
Other studies of decoherence on quantum computers have been presented 
which are not directly related to the issue of factoring. 
For example, the importance of losses and decoherence have been analyzed 
\refto{CLPY95} for the optical quantum computer designed by 
Chuang and Yamamoto \refto{ChuangYamam94} to 
solve Deuscht's Oracle problem \refto{DeutschOracle} for
a one bit function. The effect of decoherence upon a static quantum 
computer was also analyzed in \Ref{Unruh94,BarencoDeco}.  

The paper is divided in 
two parts: We first present an array of reversible
quantum logic gates which implements Shor's algorithm for factoring 
integer numbers in a quantum computer. 
To do that we first created subcomponents which 
 some specific tasks such as controlled
multiplication, controlled sums, mod(N), etc. Then, we combined 
these subcomponents in the precise way required to run Shor's algorithm. 
The existence of work--qubits (required to handle the reversible logic) 
makes the design of the quantum circuit a rather nontrivial task. 
In fact, for the quantum computer to work properly, it is necessary to 
reversibly erase 
the records created along the computational path 
(stored in the work--qubits). 
As an example, we present the gate array that could be used to 
factor $N=15$ in a quantum computer. 

Designing the factoring circuit is the first step required for studying the 
impact of decoherence and the possibility of implementing 
error correction schemes. This is the 
purpose of the second part of the paper where we 
study how the coupling to an environment affects the 
functioning of the quantum factoring circuit. 
For this, we use an oversimplified model of the system--environment 
interaction. We assume that this interaction takes place only at 
certain (randomly chosen) moments of time affecting only 
a few (randomly chosen) qubits which may spontaneously decay.

After completing the 
design of the factoring circuit, and while we were 
working on the numerical simulations to model dissipation, 
we became aware that a very similar gate array was recently 
developed by Vedral, Barenco and Ekert \refto{OxfordFact95}. Our circuit 
produces the same final quantum state and has roughly the same requirements 
(in number of qubits and time steps) than the one described in 
\Ref{OxfordFact95} (in that paper the authors did not attempt to 
analyze the impact of losses and decoherence on the performance of their
quantum circuit, an issue which we analyze here).  
More recently Plenio and Knight \refto{Knight95} used some of the conclusions 
of \Ref{OxfordFact95} (the number of required qubits and time steps) to 
discuss some of the limitations imposed by decoherence on the size of 
the numbers one could factorize using various physical setups.

In Section 2 we briefly describe both the mathematical basis for Shor's 
algorithm and the basic steps a quantum computer would need to follow 
in order to implement it. In Section 3 we describe the 
quantum network for implementing modular exponentiation. 
We go from the coarser description where the 
circuit is just a black box to the fine grained picture where every 
component is dissected and built from elementary Toffoli gates. We analyze
the architecture required to factor numbers of $L$ bits and explicitly   
exhibit the circuit to factor $N=15$, which requires $28$ qubits (the 
circuit to factor $L$ bit numbers needs $5L+8$ qubits and involves 
a number of elementary gates which, for large $L$ is close to $240 L^3$). 
In Section 4 we address the importance of decoherence 
and the possible strategies for error correction. We 
summarize our results in Section 5.

\subhead{2. Shor's algorithm}

In 1994, Peter Shor invented an algorithm for a quantum computer that 
could be used to find the prime factors of integer numbers in 
time. We will now briefly review the most important aspects of Shor's 
algorithm and later consider the way to implement it in a 
quantum computer.

The mathematical basis for Shor's algorithm is the following:
(see \Ref{Shor1, Shor2, EckertJosza95}): 
The goal is to find the prime factors of an integer 
number $N$. Instead of doing this directly, the algorithm finds the {\it 
order}, $r$, of a number $x$. The order of $x$ is 
defined as the least integer $r$ such that $x^r\equiv 1$ (mod$N$). Knowing
$r$ one can find the prime factors of $N$ by using some results proved
in Number Theory. Factorization reduces to 
finding $r$ if one uses a randomized algorithm: 
as Shor shows in \Ref{Shor2}, choosing 
$x$ at random and finding its order $r$, one can find a nontrivial 
factor by computing the greatest common
divisor $a=$gcd$(x^{r/2}-1,N)$. In fact, $a$ is a nontrivial 
factor of $N$ unless $r$ is odd or $x^{r/2}=-1$ mod$N$. As $x$ is 
chosen at random, the probability for the method yielding a nontrivial 
prime factor of $N$ is $1-1/2^{k-1}$, where $k$ is the number of 
distinct prime factors of $N$. 

In his seminal paper \refto{Shor1, Shor2}, 
Shor showed that a quantum computer could efficiently find the order $r$ 
of the number $x$ and, therefore, factorize $N$ in polynomial time. 
Let us now describe the basic operation of this quantum computer. 
This requires two quantum registers, which hold integers
represented in binary notation. There should also be a number of 
work--qubits, which are required along the calculation but should be in a 
standard state (say $|0\rangle $) both at the beginning and at the end of 
the calculation. The role of these work--qubits is very important and 
will be described in detail in the next section. For the moment, we
will concentrate on describing the state of the computer before and 
after every major step of the program. 
For that purpose, we can forget for the moment these qubits.
Apart from the quantum registers, 
there is also some classical information we should provide for 
operating the quantum computer. Thus, we will assume that the numbers
$N$ (the one we want to factor), $x$ (chosen randomly mod$N$) and a 
randomly chosen $q$, which is such that $N^2\leq q\leq 2N^2$ are 
part of the classical information available to the quantum computer. 

We start the process by preparing the first register in a uniform superposition
of the states representing all numbers $a\leq q-1$ (this can be done 
by a standard technique, i.e. rotating each individual qubit putting it
in a superposition ${1\over \sqrt 2}(|0\rangle +|1\rangle )$). The state of the 
computer is then
$$
|\Psi_0\rangle ={1\over \sqrt q} \sum_{a=0}^{q-1} \ |a\rangle \ |0\rangle \eqno(psi0)
$$
The next step is to unitarily evolve the computer into the state
$$
|\Psi_1\rangle ={1\over 
\sqrt q} \sum_{a=0}^{q-1} \ |a\rangle \ |x^a\ ({\rm mod}N)\rangle .\eqno(psi1) 
$$

Then, we Fourier transform the first register. That is, we
apply a unitary operator that maps the state $|\Psi_1\rangle $ into
$$
|\Psi_2\rangle ={1\over q} \sum_{a=0}^{q-1} \sum_{c=0}^{q-1}
 \exp(2\pi i a c/q) \ |c\rangle \ |x^a \ ({\rm mod}N)\rangle . \eqno(psi2)
$$

The final step is to observe both registers (the method
could be implemented observing just the first register but, following Shor
\refto{Shor2}, for clarity we assume both registers are observed). 
The probability for finding the state $|c\rangle \ |x^k \ ({\rm mod}N)\rangle $ is:
$$
P\bigl(c, x^k\ ({\rm mod}N)\bigr)=\Bigl|{1\over q}\sum_{a/x^a\equiv x^k}
\exp(2\pi i a c/q)\Bigr|^2, \eqno(probc1)
$$
where the sum is over all numbers $0\leq a\leq q-1$ such that 
$x^a=x^k ({\rm mod}N)$. This sum can be transformed into
$$
P\bigl(c, x^k\ ({\rm mod}N)\bigr)=\Bigl|{1\over q}\sum_{b=0}^{[(q-1-k)/r]}
\exp(2\pi i b \{rc\}_q/q)\Bigr|^2, \eqno(probc2)
$$
where $\{rc\}_q$ is an integer in the interval $-q/2< \{rc\}_q\leq q/2$
which is congruent to $rc$ (mod$q$). As shown by Shor, the above probability
has well defined peaks if $\{rc\}_q$ is small (less than $r$), i.e., if
$rc$ is a multiple of $q$ ($rc=dq$ for some $d<N$). 
Thus, knowing $q$ and the fact that the position of the peaks $c$ will be
close to numbers of the form $dq/r$, we can find the order $r$ (using 
well established continuous fraction techniques). 

There is no doubt that Shor's algorithm would work if a quantum computer
could be built. However, to implement Shor's algorithm in a quantum computer
one needs to explicitly construct the program. 
The procedure for Fourier transforming is well known and has been 
extensively discussed in several recent papers (see \Ref{Shor2, Copper94, 
EckertJosza95}). To explicitly 
construct the unitary evolution that takes the state $|\Psi_0\rangle $ into the 
state $|\Psi_1\rangle $ is a rather nontrivial task which we will describe in 
the next section\refto{OxfordFact95}. 

\subhead{3. Quantum network for modular exponentiation.}

We will present an array of quantum gates which maps the state
$|a\rangle \otimes |0\rangle $ into $|a\rangle \otimes |x^a\ ({\rm mod}N)\rangle $ 
transforming the state $|\Psi_0\rangle $ into $|\Psi_1\rangle $. We describe 
the quantum circuit using diagrams such as the one in Figure 1 which 
must be interpreted as representing the time evolution of the system
with time flowing from left to right. 
Each line represents a single qubit, i.e. a two level system 
(a thick line will represent a bundle of qubits). 
In describing the circuit we will go in steps 
from the coarse description of Figure 1a (where the computer
is a black box) to a fine grained description where the computer 
consists of a complex array of interconnected 
elementary gates. 

We will use Toffoli gates as ``elementary'' components and follow the 
notation of \Ref{Barencoetal95} denoting a gate acting on three 
qubits as $\Lambda_2$. 
The action of a Toffoli gate on a computational state $|x_1,x_2,x_3\rangle $ (where 
$x_i\in\{0,1\}$) 
is $\Lambda_2|x_1,x_2,x_3\rangle =|x_1,x_2,x_3\oplus(x_1\wedge x_2)\rangle $ where $\oplus$ 
denotes the exclusive OR and $\wedge$ the AND operation between the Boolean 
variables $x_i$.
Thus, Toffoli gates are just controlled--NOT gates where the last qubit changes
its state only if the two control qubits are set to $1$. 
It will also be convenient to use generalized Toffoli gates, with $n$ 
control--qubits, which are denoted as $\Lambda_n$. 
Of course, all these gates can be constructed in terms of one and 
two--qubit operations, as explained in \Ref{Barencoetal95}.  
The diagram representing the gate $\Lambda_n$ is shown in Figure 1b.

To design a quantum circuit for modular exponentiation 
we should first notice that if the binary representation 
of $a$ is $a=\sum_{i=0}^na_i 2^i$, then
$$
y^a\ ({\rm mod}N) = 
\prod_{i=0}^n \Bigl(\bigl(y^{2^i}\bigr)^{a_i}\ ({\rm mod}N)\Bigr). \eqno(ytotheamodn)
$$
Thus, modular exponentiation is just a chain of products where each 
factor is either equal to $1$ if $a_i=0$ or equal to $y^{2^i}$ if 
$a_i=1$. Therefore, the circuit 
is easily constructed if one is allowed to use a controlled
multiplier as an auxiliary unit (which at this level, acts as 
a new black box). In Figure 2 we show the basic architecture of the array
of controlled multipliers required for modular exponentiation. 
For the first multiplication the control qubit is $a_0$ and after 
each multiplication the control is moved to the next qubit. 
For this array to work we need to know all the numerical factors entering in
\(ytotheamodn) 
(thus, we must classically compute the numbers $y^{2^i} \ ({\rm mod}N)$).

Our next step is to analyze the controlled multiplier. 
Given an input $|I\rangle $, this circuit, which we denote as $\Pi_N(C)$,
produces an output $|I*C ({\rm mod}N)\rangle $. 
The controlled multiplier is constructed using 
a smaller black box: a controlled mod$N$ adder. 
In fact, multiplication of two numbers $I=\sum_{i=0}^nI_i2^i$ and $C$
reduces to a sum of the form $\sum_{I=0}^n I_i*\bigl(2^i C\bigr)$. 
Thus, we just need to use $I_i$ as the control qubit in a controlled 
mod$N$ adder adding the number $(2^iC)$ (a circuit which we denote
as $S_N(2^iC)$). The numbers involved in the 
sum must also be provided as classical information (we need 
to classically compute all numbers $2^jy^{2^i}$, with $i,j\leq L$ where $L$ 
is the number of bits of $N$). In Figure 3 we show a controlled 
multiplier for $4$--bit numbers. The same architecture can be used 
to multiply $L$--bit numbers. 
In that case, the controlled multiplier requires 
$L+1$ work--qubits, whose state 
is set to zero before and after its operation. 
As we will see below, the controlled adder itself also requires
some work--space which must be independent of the one used specifically
for multiplication.

As shown in Figure 3, $\Pi_N(C)$ is schematically divided into 
three pieces. In all of them the work--qubits play an important role.
The quantum state entering the circuit is 
$|\chi_0\rangle =|I\rangle \otimes|0\rangle _{wb}$, where $I$ is the number stored in the input 
register and $|0\rangle _{wb}$ is the state of the work--qubits. The qubits 
$|I_i\rangle $ are used as control for the $S_N(2^iC\ {\rm mod}N)$  
adders and the result of the sum is temporarily written in 
the work--qubits. After this, the state is $|\chi_1\rangle =|I\rangle \otimes|I*C\rangle _{wb}$: 
almost what we need, except for the fact that the input $|I\rangle $ also 
appears in the output state.  Erasing this  
extra copy of the input is essential:  
Otherwise we would be keeping a record of the computational path 
affecting the interference pattern of the quantum computer (appart
from forcing us to use an enormous ammount of space). The reversible erasure
of the input is the purpose of the second part of the circuit. 
In designing this we followed well known techniques developed by Bennett
\refto{Bennetterase} and described by Shor \refto{Shor2}. The 
procedure is as follows: We first consider the evolution operator $\tilde U$ 
mapping the input $|0\rangle \otimes|I'\rangle _{wb}$ into $|I'*C^{-1}\rangle \otimes|I'\rangle _{wb}$, 
where $C^{-1}$ is the multiplicative inverse of $C$ (mod$N$) (the number 
satisfying $C*C^{-1}=1$ (mod$N$)). The operator needed in the second 
part of the multiplier is $\tilde U^{-1}$. 
To convince ourselves that this is the case, we should 
notice that, as the input to the second part of the multiplier is 
$|\chi_1\rangle =|I\rangle \otimes|I*C\rangle _{wb}$, the output will 
be $|\chi_2\rangle =\tilde U^{-1}|\chi_1\rangle =|0\rangle \otimes |I*C\rangle _{wb}$ 
(because, by construction, $\tilde U$ satisfies $\tilde U\bigl(|0\rangle \otimes|I*C\rangle _{wb}\bigr)=|I\rangle \otimes|I*C\rangle _{wb}=|\chi_1\rangle $). 
The circuit for $\tilde U^{-1}$, shown in the figure, 
is just the specular image of the one 
used for the first part of the multiplier (switching the role of 
register and work--qubits). Finally, the multiplier is 
completed with a controlled
swap that interchanges once more the register and work--qubits so that the 
final state of the work--qubits is always $|0\rangle _{wb}$. 

The circuit for doing controlled mod$N$ sums of a number $X$, which is 
stored in a quantum register, and a number $Y$, stored in a classical register, 
is called $S_N(Y)$. This circuit, for $5$--bit numbers, 
is shown in Figure 4 (generalization to 
$L$ bit numbers is straighforward). The circuit for $S_N(Y)$ is built
using a simple controlled adder, which we denote as $S(Y)$ 
whose functioning will be explained below. The only 
difference between $S_N(Y)$ and $S(Y)$ is that the former 
gives the output modulo N. 
Constructing a reversible circuit for computing the sum mod$N$ is not a trival
task which is only possible because we know that the two numbers being 
added ($X$ and $Y$) are both less than $N$ (and therefore $X+Y\leq 2N-2$). 
Without this information it would not be possible to compute mod$N$ 
reversibly without keeping unwanted records of the computation (since 
mod$N$ is not a one to one function). 
The input to the circuit is $|\bar\chi_0\rangle =|X\rangle \otimes|0\rangle _{wb}$. 
After the first adder, this is transformed to 
$|\bar\chi_1\rangle =|X+Y\rangle \otimes|0\rangle _{wb}$. We then apply another simple adder 
adding the possitive number $2^{L+1}-N$ 
producing an output $|\bar\chi_2\rangle =|2^{L+1}+X+Y-N\rangle \otimes|0\rangle _{wb}$. 
The most significant 
bit (MSB) of $2^{L+1}+X+Y-N$ is one (zero) if $X+Y\geq N$ ($X+Y<N$). It is 
easy to realize that the opposite 
is true for the second MSB of the output. 
Thus if we use this qubit to control
the inverse operation, we will add $N$ only if $X+Y<N$. 
Therefore, after the third gate of the circuit shown in Figure 4, 
the first $L$ qubits of the output always store the number $A+C$ mod$N$. 
However, the $L+1$-- and $L+2$--qubits, which are used to control the 
third gate, keep a record of the first result. As usual, this record 
must be reversibly erased and this can be done by using the following simple
trick: We first add the possitive number $2^{L}-Y$ and notice 
that the MSB of the result $2^{L}-Y+(X\ {\rm mod}N)$ is always identical to the 
qubit used to control the third gate. Thus, we are done: we apply a 
control--NOT gate and then we undo the first sum (by adding $Y$). 

So far, we first explained modular exponentiation in terms of 
controlled multiplication $\Pi_N(C)$. 
Later, we explained $\Pi_N(C)$ 
in terms of controlled mod$N$ sums $S_N(Y)$) and this 
circuit in terms of a simple adder $S(Y)$. 
We will now present
the gate array for the simple controlled adder $S(X)$ which is best 
explained in terms of a smaller gate: a controlled two--qubit adder. 
This will be our smallest black box and, for clarity, we will explain 
here how it works. The two--qubit 
adder, denoted as $\Sigma(\sigma)$ has four input qubits and a classical 
input bit $\sigma$ (i.e., there are two types of two--qubit adders, one 
for $\sigma=0$ and another for $\sigma=1$). The first input qubit is 
the control, the second qubit is $i_1$, the third one is $i_2$ and the fourth one is a work--qubit which is always set to $0$ at the input. At the 
output, the control qubit 
is unchanged, the first qubit changes into the Least Significant Bit (LSB)
of the sum ($i_1+i_2+\sigma$), the third one stores $i_2$ and the 
fourth stores the MSB of the sum. In Figure 5 we
can see how to build the gates $\Sigma(0)$ and $\Sigma(1)$ (and other 
useful simple gates) in terms of Toffoli gates.

Using $\Sigma(\sigma)$ it is possible to 
construct a circuit mapping an input $|X\rangle $ into 
$|X+Y\rangle $. This is displayed in Figure 6 where, for simplicity, 
we assumed that both $X$ and $Y$ have $5$ bits. For  
numbers of $L$ bits the number of work--qubits required is $L+3$. 
The quantum state entering the adder 
is $|\tilde\chi_0\rangle =|X\rangle \otimes|0\rangle _{wb}$. This goes through the  
sequence of two--qubit adders $\Sigma(Y_i)$ (we use $X_i, Y_i\in\{0,1\}$ 
for the binary representation of $X$ and $Y$). 
After this chain of $\Sigma$--gates, the state is  $|\tilde\chi_1\rangle =|X\rangle \otimes|X+Y\rangle _{wb}$, which has an unwanted copy 
of the input. To reversibly 
erase this extra copy we apply the same method used in the 
multiplication: We first consider an auxiliary operator $W$ that adds
the possitive number $\bar Y\equiv 2^L-Y$ ($\bar Y$ is 
known as the two's complement of $Y$ and its 
binary representation is simply obtained from that of $Y$ by 
interchanging zeros and ones and adding $1$). The operator $W$ 
satisfies $W\bigl(|R\rangle \otimes|0\rangle _{wb}\bigr)=|R\rangle \otimes|R+2^L-Y\rangle _{wb}$. 
Therefore, its inverse is such that 
$W^{-1}|X+Y\rangle \otimes|2^L+X\rangle _{wb}=|X+Y\rangle \otimes|0\rangle _{wb}$ which is precisely 
what we need as the output of our circuit 
(the properties of $W^{-1}$ simply follow from  that of $W$ which, 
by construction satisfies 
$W|X+Y\rangle \otimes|0\rangle _{wb}=|X+Y\rangle \otimes|2^L+X\rangle _{wb}$). 
Therefore, using $W^{-1}$ after appropriately interchanging the role of 
the register and the work--qubits (and adding an extra work--qubit 
to store the qubit representing $2^L$) we complete the controlled adder. 
The circuit for $W^{-1}$ which is shown in Figure 6 is almost the 
specular image from the one used as the first part of the adder. The 
only difference is that instead of the first two qubit adder we can 
use a smaller circuit which only stores the LSB of the first sum (this 
circuit is shown in Figure 5). 
\bigskip

Having explained the essencial pieces of the quantum computer, let us
now summarize what are its space and time requirements 
(i.e., the number of qubits and the number of elementary operations). 
As explained above, to factor an $L$ bit number we need:
$L+1$ qubits as work--space for the controlled multiplier and $L+4$ for
controlled sums. The mod$N$ circuit as well as the controlled swap require 
an extra work--qubit each. Adding 
the qubits required to store the two quantum registers ($2L+1$ qubits to 
store $a$ in the first register and $L$ qubits for the second register) 
we get a total of $5L+8$ qubits. Computing the number of 
elementary operations is also possible. By inspecting our 
controlled adder one realizes that the number of elementary 
gates is $\alpha L+\beta (L+1) +(L+2) \gamma$ where $\alpha$, $\beta$ 
and $\gamma$ are, respectively, the number of gates in a two--qubit adder, 
its inverse and the one in a swap circuit. Using the 
estimate $\alpha=\beta=3$ one gets $12 n +17$ operations for the sum. 
Using similar arguments to analyze the multipliers one finally 
concludes that the complete modular exponentiation circuit requires
$240 n^3 + 484 n^2 + 182 n$ elementary operations. For $L=4$ this is
about $2.5 \ 10^4$.

\subhead{4. Losses and decoherence in a factoring computer}

Before analyzing the impact of dissipative effects on the 
quantum circuit it is convenient to introduce some notation. 
The quantum computer has a Hilbert space with a computational basis with 
states $|r_1, r_2, wb\rangle $ (where $r_1$, $r_2$ and 
$wb$ are the bit--strings determining the states of the first register, 
the second register and the work--qubits respectively). We assume 
that the environment ${\cal E}$ has 
a Hilbert space spanned by a basis of states $|e\rangle _{\cal E}$. 
The quantum state of the computer--environment ensemble 
can always be written as
$$
|\Psi(t)\rangle =\sum_{r_1, r_2, wb, e} 
A(r_1, r_2, wb, e, t) \ |r_1, r_2, wb\rangle  |e\rangle _{\cal E}. 
\eqno(statet)
$$
The temporal evolution of the 
probability amplitude $A(r_1, r_2, wb, e, t)$ is governed by 
the interplay between the quantum circuit described in Section 3 and 
the computer--environment interaction. At the initial time, when the computer 
is in state \(psi0), the amplitudes are given by:
$$
A(r_1, r_2, wb, e, t=0)={1\over \sqrt q}
\delta(r_2,0) \ \delta(wb,0)\  \delta(e,0).\eqno(amplit0)
$$  
Here we assumed that the computer is initially 
uncorrellated with the environment which is taken to be in an unexcited 
state $|0\rangle _{\cal E}$ (we use $\delta(a,b)$ to denote 
Kronecker's delta function). 
If the computer evolves without interaction with the 
environment the amplitudes after the modular exponentiation
circuit are:
$$
A_{exact}(r_1, r_2, wb, e, t=t_f)={1\over \sqrt q} 
\delta(r_2,y^{r_1}(mod N))\ \delta(wb,0)\  \delta(e,0).
\eqno(amplittf)
$$  

However, when the computer interacts with the environment, the actual 
amplitudes will deviate from the exact expression \(amplittf). To model 
this interaction we will use a very simple approach which 
incorporates the losses induced by the spontaneous decay 
of the computer's qubits: The environment consists of a collection of 
two level systems ${\cal E}_i$, i.e. a collection of 
``environmental qubits'' (each ${\cal E}_i$--qubit has an excited 
state $|1\rangle _{{\cal E}_i}$ and a ground state $|0\rangle _{{\cal E}_i}$). 
For simplicity we will assume that at a given time, a randomly 
selected computer qubit $q_i$ interacts with one of environmental qubits
${\cal E}_i$. As a result of this sudden interaction correlations 
are established according to:
$${\eqalign{
|1\rangle _{q_i}\ |0\rangle _{{\cal E}_i}\rightarrow& 
p_1^{1/2}\ |1\rangle _{q_i}\ |0\rangle _{{\cal E}_i}+p_2^{1/2}\ |0\rangle _{q_i}
\ |1\rangle _{{\cal E}_i}\cr
|0\rangle _{q_i}\ |0\rangle _{{\cal E}_i}\rightarrow & |0\rangle _{q_i}
\ |0\rangle _{{\cal E}_i}\cr}}\eqno(trules)
$$
where $p_2=1-p_1$. The interpretation of the evolution \(trules) is quite 
clear: If the computer qubit is in the state $|1\rangle _{q_i}$ it has a 
probability $p_1$ to persist and a probability $p_2$ to decay into 
$|0\rangle _{q_i}$ creating an excitation in the environment. On the other 
hand, if the computer qubit is in the state $|0\rangle _{q_i}$ nothing happens.
It is worth mentioning that the decay rules \(trules) implicitly assume that 
the state used to represent the {\it computational} $0$ is the ground state
(or, at least, has lower energy than the one used to represent the 
{\it computational} $1$). In fact, the situation may be exactly the opposite in 
which case the rules \(trules) must be trivially modified by interchanging 
the roles of $|1\rangle _{q_i}$ and $|0\rangle _{q_i}$ (see below). 
More general evolution rules (such as the ones used in \Ref{Shor95}, 
which are best suited to analyze a noisy but almost losseless computer) 
will be studied elsewhere \refto{us96}. 

Thus, we can summarize the basic ingredients of our 
computer--environment model: i) It is caracterized by 
a randomly chosen sequence of times $(t_1,\ldots, t_n)$ which define
the instants where the computer interacts with the environment (in 
between these times the computer evolves according to the unitary 
operators associated with the quantum circuit described in the previous
section). ii) At each time $t_i$ we randomly choose a computer qubit $q_i$ 
which is involved in a sudden interaction 
with an environmental qubit ${\cal E}_i$. iii) As a consequence of 
this interaction the computer--environment ensemble 
evolves according to the 
rules \(trules). 
Implicit in our assumptions is the validity of the 
simplifying Markovian approximation which assures 
that at every instant $t_i$ a different (and independent) 
environmental qubit ${\cal E}_i$ is involved in the interaction. 
A simple way of visualizing this computer--environment model is 
by thinking of the times $t_i$ as the instants where there may be a 
``branching'' of the computational trajectory. Every time an environmental 
qubit is excited an ``erroneous'' computational trajectory emerges. At 
the end of the modular exponentiation circuit, 
the state vector of the computer--environment
ensemble is written as in \(statet) with an amplitude which will 
{\it not} be given by \(amplittf). 
We already admitted that this is an oversimplification of reality (which 
has been used before to model losses in quantum computation 
\refto{CLPY95}).

We computed the amplitudes from the output state of 
the Fourier Transform circuit which follows modular 
exponentiation (the discrete FT circuit is described in the literature
\refto{Shor2, CiracZoller95, Copper94}). In Figure 7 
results are presented for the probability
for finding $r_1$ in the first register and $r_2=7$ in the 
second register. The ideal result, ploted in Fig. 7(a), is obtained from 
eq. \(probc2). 
This error--free curve has three sharp peaks, with a separation approximately 
equal to $q/r=130/4$ (we deliberately choose a rather small value for $q$ so that the 
small structure in the plots can be seen using a reasonable scale).  
Provided we don't know the final state of the environment and the 
work--qubits (see below) the probability is
$$
P_{NED}(r_1, r_2)=\sum_{wb, e} \bigl|A(r_1, r_2, wb, e, t)\bigr|^2.\eqno(probne)
$$ 
(the suffix stands for ``no error detection'', see below).
This probability is shown in Figure 7(b) where we can see that the 
errors slightly widen the peaks and notably decrease their
amplitudes. As the number of errors is increased it will be 
less and less likely to measure a value of $r_1$ located near
a peak making the identification of the order $r$ (obtained from the
separation between peaks, as explained in \Ref{Shor2}) more and more difficult.
The appearence of intermediate peaks is also evident in Fig. 7(b).
Appart from the above probability 
we also calculated the probability for finding $r_1$ in the first 
register, $r_2=7$ in the second and the work--qubits 
in the state $|0\rangle _{wb}$, i.e.: 
$$
P_{ED}(r_1, r_2)=\sum_{e} \bigl|A(r_1, r_2, wb=0, e, t)\bigr|^2.\eqno(probec)
$$ 
This is plotted in Fig. 7(c) where we see that while a noisy dc 
component (present 
in (b)) is supressed, the amplitude ratio between the misleading and correct
peaks is increased. These plots correspond to  simulations of the quantum 
computer running the program to factor $N=15$ while 
coupled to an environment at a randomly chosen set of ten instants $t_i$
(we use $p_1=p_2=1/2$). The modular exponentiation circuit requires about
$2.5\  10^{4}$ elementary (Toffoli) gates. This roughly 
correspond to $10^{5}$ one bit operations for Cirac and Zoller's 
cold ions computer \refto{CiracZoller95}, Thus, in that case 
we are considering an error rate of the order 
of $10^{-4}$, which is a rather optimistic figure. 

Our simulations not only can be used to visualize the 
importance of the environmental interaction on the quantum algorithm but 
also to test simple error detection (and correction) schemes. 
The simplest of such schemes is probably the one based on checking 
the state of the qubits which are supposed to be in a known state. 
Our factoring program is suited for this purpose since the work--qubits 
must start and end in the state representing the computational $0$.  
Two comments concerning error detection (and correction) are in order:
First, by checking the final state of the work--qubits we are not able to 
detect a special class of errors which are produced by the decay of 
the qubits representing the first and second registers of the computer ($r_1$ 
and $r_2$). Errors of that kind leave (most of the time) 
the work--qubits untouched but generate a misleading output (they are 
responsible for the intermediate peaks seen in Figure 7(c) which 
make the measurement of the order $r$ a much more difficult task).  
Second, and more important, by measuring the final state of the work--qubits we 
are only able to the {\it detect} errors but not to correct (or prevent) 
them. 

Of course, it would be much better to have a method
enabling us to {\it prevent} the errors from occuring.  For this, the 
use of the watchdog effect \refto{watchdog} has been proposed. 
Thus, if some of the computer's qubits are  
supposed to be in a known state at some time, one could inhibit their decay by 
making a measurement on the known state. This method can 
indeed be applied here since the work--qubits are supposed to be 
in the state representing the computational $0$ 
at many intermediate instants of the computation. In fact, this is what 
happens after the action of each $\Pi_N(C)$ circuit and after the action 
of each 
controlled adder $S_N(C)$. For large $L$, the number of times one could 
measure the state of some of the work--qubits grows as $L^2$. 

To test the efficiency of the watchdog effect as an error correction
technique we slightly changed our computer--environment interaction model. 
In fact, we now assume that the decay rules are of the form \(trules) but 
with time dependent coefficients given by:
$$
p_1(t)=\exp(-\gamma t), \ \ p_2(t)=1-p_1(t).\eqno(timedepprob)
$$
In this way the decay probability for a qubit increases with time (measured 
from the start of the computation and, by convention, expressed  
in units of the total time required to run the program, i.e. $t=1$ corresponds
to the end of the computation). The decay rate $\gamma$ is taken to be $\gamma=2.5$ 
so that towards the end of the computation a qubit will have a high decay probability
($p_2(t=1)\approx 9/10$). The assumption of an exponential decay is not 
essential (it is just a reasonable approximation which we addopt here for 
simplicity). 

To implement the watchdog we measure
the state of the work--qubits at every instant when they are supposed to be
in the computational $0$. Every time we do this we reset the time in 
\(timedepprob). Thus, a work--qubit will decay with probabilities given by 
\(timedepprob) where the time will effectively be measured from the last 
instant 
in which the work--qubit was supposed to be in the computational $0$ state. 
On the other hand, the qubits involved in the first or second registers of the 
computer will have decay probabilities given by \(timedepprob) with time counting
from the begining of the computation. 

The effectiveness of the watchdog effect as an error prevention 
technique can be seen in Figure 8 where the 
exact probability is plotted together with the ones obtained with and without watchdog. Without using this method we 
get a very noisy probability with a substantial widening of the principal 
peaks. The amplitude of the central peak, which is about $0.1$, is
of the same order as 
the one shown in Fig. 7(b) (but the decay rules we are using 
here are more damaging than the ones we used before). Using the watchdog
technique we substantially increase 
the amplitude of the main peaks (by a factor of four)
and also eliminate almost all the noise. 
The only remaining spurious peaks are those produced by
the decay of qubits involved in the first and second registers. They can not be 
eliminated using watchdog since their existence is not a consequence of a 
process affecting the work--qubits.

\subhead{5. Summary and outlook.}

The factoring circuit we presented is by no means optimal. Several 
improvements are possible to reduce the number of work--qubits. However, 
when designing a circuit for practical purposes one has to have in mind 
that the existence of work--qubits is not necesarily a burden. 
Our results show they can play a very useful role allowing the use of the 
watchdog effect as an error prevention technique. It would be important to
find the optimal balance minimizing the number of work--qubits but still 
allowing an efficient use of the watchdog method. 

The simulations we performed are rather simple and do not allow us to test the importance of other sources of problems for quantum computers. 
One of the most important sources of errors we excluded here is related with
the fact that the elementary quantum gates are 
never $100\%$ efficient. If we think of Cirac and Zoller's
\refto{CiracZoller95} cold ions hardware, the elementary gates are built
by applying a sequence of laser pulses on individual ions. If these 
pulses are not exact $\pi$--pulses (or $\pi/2$--pulses) the 
quantum gate will not be exactly the one we want. The corresponding 
unitary evolution operator $U_{real}$ 
will have nonzero matrix elements in places where the exact quantum gate 
operator $U_{ideal}$ has zero matrix elements. 
These imperfections may be rather important since their effects 
accumulate in time. To include this effects in our model 
one needs to follow the evolution of the computer's 
state vector in the $2^{28}$--dimensional Hilbert space. 
Even though our work enables us 
to explicitly write down the matrix $U_{real}$ at every step 
of the calculation, we are not able to numerically simulate this because
of space limitations (thus, simulating a quantum computer with $N$ qubits
needs an exponentially large ammount of space in a classical computer). 
Simulations of smaller versions of our circuit for modular exponentiation
will be presented elsewhere \refto{us96}.

\references

\refis{Seth95} S. Lloyd, Scientific American {\bf 273}, 44 (1995).

\refis{BennettNat} C. Bennett, Physics Today, 24, October (1995);
C. Bennett and D. DiVincenzo, Nature {\bf 377}, 389 (1995). 

\refis{Shor1} P. Shor, in {\it Proceedings of the 35th Annual Symposium
on Foundations of Computer Science}, edited by S. Goldwasser (IEEE Computer Society, Los Alamitos, CA, 1994), p. 116.

\refis{Shor2} P. Shor, ``Polynomial time algorithms for prime factorization 
and discrete logarithms on a quantum computer'', preprint 
quant-ph/9507027 (1995).

\refis{Barencoetal95} A. Barenco et al. Phys. Rev. {\bf A52}, 3457 (1995). 

\refis{EckertJosza95} A. Ekert and R. Josza, ``Shor's algorithm for 
factorizing numbers'', Rev. Mod. Phys. to appear (1995). 

\refis{OxfordFact95} V. Vedral, A. Barenco and A. Ekert, ``Quantum networks for 
elementary arithmetic operations'', submitted to Phys. Rev. {\bf A}. 

\refis{Griffiths} R. Griffiths and , preprint quant-ph/ (1995).

\refis{DiVincenzo94} D. DiVincenzo, Phys. Rev. {\bf A51}, 1015 (1995).  

\refis{CiracZoller95} A. Cirac and P. Zoller, Phys. Rev. Lett. {\bf74}, 
4091 (1995). 

\refis{Weinfurter95} T. Sleator and H. Weinfurter, Phys. Rev. Lett. {\bf 74}, 4087 (1995). 

\refis{Kimble95} Q. A. Turchette, C. J. Hood, W. Lange, H. Mabuchi and 
H. J. Kimble, Phys. Rev. Lett. december (1995). 

\refis{Unruh94} W. G. Unruh, Phys. Rev. {\bf A51}, 992 (1995). 

\refis{Paz94} J. P. Paz, ``Decoherence in an evolving quantum computer'' (1995) 
unpublished. 

\refis{CLSZ95} I. Chuang, R. Laflamme, P. Shor and W. Zurek, Science {\bf 270},
1633 (1995). 

\refis{Bennetterase} C. Bennett, IBM J. Res. 
Develop. {\bf 17}, 525 (1973); SIAM J. Comput. {\bf 18}, 766 (1989).

\refis{BarencoDeco} G.M. Palma, K- A. Suominen and  A. Ekert, ``Quantum
computers and dissipation'', submitted to Proc. Roy. Soc. London (1995). 

\refis{ChuangLaf95} I. Chuang and R. Laflamme, ``Quantum error correction 
by coding'', quant-ph/9511003. 

\refis{Knight95} M. B. Plenio and P. L. Knight, ``Realistic lower bounds for 
the factorization time of large numbers on a quantum computer'', 
quant-ph/9512001. 
 
\refis{us96} C. Miquel, J. P. Paz and R. Perazzo, in preparation. 

\refis{CLPY95} I. Chuang, R. Laflamme, J. P. Paz and T. Yamamoto, 
``Decoherence in a simple quantum computer'', submitted to Phys. Rev. {\bf A},
(1995).

\refis{Bennettetal95} C. Bennett et al, ``Purification of noisy entanglement
and faithful teleportation via noisy channels'', quant-ph/9511027.

\refis{Copper94} D. Coppersmith, ``An approximate Fourier Transform useful
in quantum factoring'', IBM Research Report RC19642. 

\refis{watchdog} W. H. Zurek, Phys. Rev. Lett. {\bf53}, 391 (1984). 

\refis{ChuangYamam94} I. Chuang and Y. Yamamoto, Phys. Rev. {\bf A52}, 3489 
(1995). 

\refis{DeutschOracle} D. Deutsch and R. Josza, Proc. Roy. Soc. London 
{\bf A439}, 553 (1992). 

\refis{Shor95} P. Shor, Phys. Rev. {\bf A53}, R2493 (1995). 

\refis{ZurekPT} W. H. Zurek, Physics Today {\bf 44}, 36 (1991); 
{\it ibid} {\bf 46}, 81 (1993).

\endreferences
\vfill\eject

{ \oneandathirdspace

\noindent Figure 1: a) Black box description of the circuit for modular 
exponentiation. When $N$ has four bits one needs nine qubits to represent 
$a$ and fiveteen extra qubits to be used as workspace.
b) A $\Lambda_4$ Toffoli gate with 4-control 
bits: $x_1, x_2, x_3, x_4$. 
$x_5 \rightarrow x_5 \oplus (x_1 \wedge x_2 \wedge x_3 \wedge x_4)$

\noindent Figure 2: The gate array used for modular exponentiation. 
$Y^a mod \ N$ is calculated by repeatedly multiplying the second 
register by $Y^{2^m} mod \ N$ only if $a_m = 1$. Each box multiplies its input 
by $Y^{2^m} mod \ N$ only if the control bit $a_m$ is 1. 

\noindent Figure 3: a) The 3 stages of the controlled multiplier 
(mod N) $\Pi_N(C)$: first the input $I$ is multiplied by $C$. 
Then $I$ is reversibly erased and finally the result is swapped with 
the upper register. b) Multiplication by $C$ is achieved by repeated 
addition of $2^mC \ mod \ N$ controlled by $I_m$. This is done using 
the controlled mod N adders $S_N(2^mC \ mod \ N)$. In the Figure we denote
mod$N$ as $\%N$. 

\noindent Figure 4: Addition mod N is achieved with 5 controlled adders: 
The first adds $C$ to the input. The second ``subtracts" $N$ 
from $a+C$. The third operation adds $N$ only if $a+C$ is smaller 
than $N$. At this stage the first 4 bits have $a+C \ mod \ N$. The 
last two stages erase the record left in the 7th bit, whose state depends
on the sign of $a+C-N$.

\noindent Figure 5: 
The two-qubit adders $\Sigma( \sigma )$ are shown in terms of Toffoli 
gates. They have four input and four output qubits. If the inputs are 
$ctl$, $i_1$, $i_2$ and $0$, the outputs are 
$ctl$, the least significant bit (LSB) of $i_1+i_2+\sigma$, $i_2$ and the most 
significant bit (MSB) of the sum. 
A swap gate is also shown that interchanges its two input qubits: $i_1$ and 
$i_2$. 

\noindent
Figure 6: a) Addition is performed in 3 stages: The first adds $Y$ to the input $X$, the
second interchanges $X$ with $X+Y$ and the last reversibly erases the input $X$. b) The
first and last stages are shown in terms of the individual qubits and two-qubit adders 
$\Sigma ( \sigma )$. $Y_0 \ldots Y_4$ are the bits in the binary representation of $Y$. 
$\bar Y \equiv 2^L - Y$ is used to erase $X$.

\noindent
Figure 7: Probability distribution for $r_1$ and $r_2 = 7$. 
In the simulations $N=15$, $q=130$ and $p_1=p_2=1/2 \ \forall \ t \in [0,1]$. 
a) Exact result. b) Result with ten decaying qubits at randomly chosen 
instants of time $t_1 \ldots t_{10}$. 
c) Probability distribution for $r_1$, $r_2=7$ and all work-qubits in 
their zero state.

\noindent
Figure 8: Probability distribution for $r_1$ and $r_2=7$. 
In the simulation $N=15$, $q=130$ and the decay rate $\gamma$ is chosen 
so that $p_2(t=1) \approx 9/10$. a) Exact result. 
b) Result with ten decaying qubits and using the watchdog 
effect on every work--qubit. c) Result with ten decaying qubits 
without using the watchdog effect.

}

\end

%% file: jnl.tex



\font\twelverm=cmr10 scaled 1200    \font\twelvei=cmmi10 scaled 1200
\font\twelvesy=cmsy10 scaled 1200   \font\twelveex=cmex10 scaled 1200
\font\twelvebf=cmbx10 scaled 1200   \font\twelvesl=cmsl10 scaled 1200
\font\twelvett=cmtt10 scaled 1200   \font\twelveit=cmti10 scaled 1200

\skewchar\twelvei='177   \skewchar\twelvesy='60


\def\twelvepoint{\normalbaselineskip=12.4pt
  \abovedisplayskip 12.4pt plus 3pt minus 9pt
  \belowdisplayskip 12.4pt plus 3pt minus 9pt
  \abovedisplayshortskip 0pt plus 3pt
  \belowdisplayshortskip 7.2pt plus 3pt minus 4pt
  \smallskipamount=3.6pt plus1.2pt minus1.2pt
  \medskipamount=7.2pt plus2.4pt minus2.4pt
  \bigskipamount=14.4pt plus4.8pt minus4.8pt
  \def\rm{\fam0\twelverm}          \def\it{\fam\itfam\twelveit}%
  \def\sl{\fam\slfam\twelvesl}     \def\bf{\fam\bffam\twelvebf}%
  \def\mit{\fam 1}                 \def\cal{\fam 2}%
  \def\tt{\twelvett}
  \textfont0=\twelverm   \scriptfont0=\tenrm   \scriptscriptfont0=\sevenrm
  \textfont1=\twelvei    \scriptfont1=\teni    \scriptscriptfont1=\seveni
  \textfont2=\twelvesy   \scriptfont2=\tensy   \scriptscriptfont2=\sevensy
  \textfont3=\twelveex   \scriptfont3=\twelveex  \scriptscriptfont3=\twelveex
  \textfont\itfam=\twelveit
  \textfont\slfam=\twelvesl
  \textfont\bffam=\twelvebf \scriptfont\bffam=\tenbf
  \scriptscriptfont\bffam=\sevenbf
  \normalbaselines\rm}



\def\beginlinemode{\endmode
  \begingroup\parskip=0pt \obeylines\def\\{\par}\def\endmode{\par\endgroup}}
\def\beginparmode{\endmode
  \begingroup \def\endmode{\par\endgroup}}
\let\endmode=\par
{\obeylines\gdef\
{}}
\def\singlespace{\baselineskip=\normalbaselineskip}
\def\oneandathirdspace{\baselineskip=\normalbaselineskip
  \multiply\baselineskip by 4 \divide\baselineskip by 3}
\def\oneandahalfspace{\baselineskip=\normalbaselineskip
  \multiply\baselineskip by 3 \divide\baselineskip by 2}
\def\doublespace{\baselineskip=\normalbaselineskip \multiply\baselineskip by 2}

\newcount\firstpageno
\firstpageno=2
\footline={\ifnum\pageno<\firstpageno{\hfil}%
\else{\hfil\twelverm\folio\hfil}\fi}
\let\rawfootnote=\footnote              
\def\footnote#1#2{{\rm\singlespace\parindent=0pt\rawfootnote{#1}{#2}}}
\def\raggedcenter{\leftskip=4em plus 12em \rightskip=\leftskip
  \parindent=0pt \parfillskip=0pt \spaceskip=.3333em \xspaceskip=.5em
  \pretolerance=9999 \tolerance=9999
  \hyphenpenalty=9999 \exhyphenpenalty=9999 }
\def\dateline{\rightline{\ifcase\month\or
  January\or February\or March\or April\or May\or June\or
  July\or August\or September\or October\or November\or December\fi
  \space\number\year}}
\def\received{\vskip 3pt plus 0.2fill
 \centerline{\sl (Received\space\ifcase\month\or
  January\or February\or March\or April\or May\or June\or
  July\or August\or September\or October\or November\or December\fi
  \qquad, \number\year)}}


\hsize=6.5truein
\vsize=8.9truein
\parskip=\medskipamount
\twelvepoint            
\doublespace            
\overfullrule=0pt       



\def\title                      
  {\null\vskip 3pt plus 0.2fill
   \beginlinemode \doublespace \raggedcenter \bf}

\def\author                     
  {\vskip 3pt plus 0.2fill \beginlinemode
   \singlespace \raggedcenter}

\def\affil                      
  {\vskip 3pt plus 0.1fill \beginlinemode
   \oneandahalfspace \raggedcenter \sl}

\def\abstract                   
  {\vskip 3pt plus 0.3fill \beginparmode
   \doublespace \narrower ABSTRACT: }

\def\endtitlepage               
  {\endpage                     
   \body}

\def\body                       
  {\beginparmode}               

\def\subhead#1{                 
  \vskip 0.25truein             
  {\raggedcenter #1 \par}
   \nobreak\vskip 0.25truein\nobreak}

\def\refto#1{$[{#1}]$}           

\def\references                 
  {\subhead{References}         
   \beginparmode
   \frenchspacing \parindent=0pt \leftskip=1truecm
   \parskip=8pt plus 3pt \everypar{\hangindent=\parindent}}

\gdef\refis#1{\indent\hbox to 0pt{\hss#1.~}}    

\gdef\journal#1, #2, #3, 1#4#5#6{               
    {\sl #1~}{\bf #2}, #3, (1#4#5#6)}           

\def\refstylenp{                
  \gdef\refto##1{ [##1]}                                
  \gdef\refis##1{\indent\hbox to 0pt{\hss##1)~}}        
  \gdef\journal##1, ##2, ##3, ##4 {                     
     {\sl ##1~}{\bf ##2~}(##3) ##4 }}

\def\refstyleprnp{              
  \gdef\refto##1{ [##1]}                                
  \gdef\refis##1{\indent\hbox to 0pt{\hss##1)~}}        
  \gdef\journal##1, ##2, ##3, 1##4##5##6{               
    {\sl ##1~}{\bf ##2~}(1##4##5##6) ##3}}

\def\endreferences{\body}

\def\figurecaptions             
  { \beginparmode
   \subhead{Figure Captions}
}

\def\endpage                    
  {\vfill\eject}

\def\endpaper                   
  {\endmode\vfill\supereject}


\def\ref#1{Ref. #1}                     
\def\Ref#1{Ref. #1}                     

\def\frac#1#2{{\textstyle{#1 \over #2}}}

\def\sla{\raise.15ex\hbox{$/$}\kern-.57em}
\def\leaderfill{\leaders\hbox to 1em{\hss.\hss}\hfill}
\def\twiddle{\lower.9ex\rlap{$\kern-.1em\scriptstyle\sim$}}
\def\bigtwiddle{\lower1.ex\rlap{$\sim$}}
\def\gtwid{\mathrel{\raise.3ex\hbox{$>$\kern-.75em\lower1ex\hbox{$\sim$}}}}
\def\ltwid{\mathrel{\raise.3ex\hbox{$<$\kern-.75em\lower1ex\hbox{$\sim$}}}}
\def\square{\kern1pt\vbox{\hrule height 1.2pt\hbox{\vrule width 1.2pt\hskip 3pt
   \vbox{\vskip 6pt}\hskip 3pt\vrule width 0.6pt}\hrule height 0.6pt}\kern1pt}

\def\m@th{\mathsurround=0pt }
\def\leftrightarrowfill{$\m@th \mathord\leftarrow \mkern-6mu
 \cleaders\hbox{$\mkern-2mu \mathord- \mkern-2mu$}\hfill
 \mkern-6mu \mathord\rightarrow$}
\def\overleftrightarrow#1{\vbox{\ialign{##\crcr
     \leftrightarrowfill\crcr\noalign{\kern-1pt\nointerlineskip}
     $\hfil\displaystyle{#1}\hfil$\crcr}}}


\font\titlefont=cmr10 scaled\magstep3

\def\martinstyletitle                      
  {\null\vskip 3pt plus 0.2fill
   \beginlinemode \doublespace \raggedcenter \titlefont}

\font\twelvesc=cmcsc10 scaled 1200

\def\author                     
  {\vskip 3pt plus 0.2fill \beginlinemode
   \singlespace \raggedcenter\twelvesc}


\def\heading                            
  {\vskip 0.5truein plus 0.1truein      
   \beginparmode \def\\{\par} \parskip=0pt \singlespace \raggedcenter}

\def\subheading                         
  {\vskip 0.25truein plus 0.1truein     
   \beginlinemode \singlespace \parskip=0pt \def\\{\par}\raggedcenter}

\def\tag#1$${\eqno(#1)$$}

\def\align#1$${\eqalign{#1}$$}

\def\aligntag#1$${\gdef\tag##1\\{&(##1)\cr}\eqalignno{#1\\}$$
  \gdef\tag##1$${\eqno(##1)$$}}

\def\endaligntag{}

\def\overset #1\to#2{{\mathop{#2}\limits^{#1}}}
\def\underset#1\to#2{{\let\next=#1\mathpalette\undersetpalette#2}}
\def\undersetpalette#1#2{\vtop{\baselineskip0pt
\ialign{$\mathsurround=0pt #1\hfil##\hfil$\crcr#2\crcr\next\crcr}}}


\def\ref#1{Ref.~#1}                     
\def\Ref#1{Ref.~#1}                     
\def\[#1]{[\cite{#1}]}
\def\cite#1{{#1}}
\def\(#1){(\call{#1})}
\def\call#1{{#1}}
\def\taghead#1{}
\def\frac#1#2{{#1 \over #2}}

\def\12{{1\over2}}

\def\sla{\raise.15ex\hbox{$/$}\kern-.57em}
\def\leaderfill{\leaders\hbox to 1em{\hss.\hss}\hfill}
\def\twiddle{\lower.9ex\rlap{$\kern-.1em\scriptstyle\sim$}}
\def\bigtwiddle{\lower1.ex\rlap{$\sim$}}
\def\gtwid{\mathrel{\raise.3ex\hbox{$>$\kern-.75em\lower1ex\hbox{$\sim$}}}}
\def\ltwid{\mathrel{\raise.3ex\hbox{$<$\kern-.75em\lower1ex\hbox{$\sim$}}}}
\def\square{\kern1pt\vbox{\hrule height 1.2pt\hbox{\vrule width 1.2pt\hskip 3pt
   \vbox{\vskip 6pt}\hskip 3pt\vrule width 0.6pt}\hrule height 0.6pt}\kern1pt}
\def\tdot#1{\mathord{\mathop{#1}\limits^{\kern2pt\ldots}}}

\def\pmb#1{\setbox0=\hbox{#1}%
  \kern-.025em\copy0\kern-\wd0
  \kern  .05em\copy0\kern-\wd0
  \kern-.025em\raise.0433em\box0 }

\catcode`@=11
\newcount\tagnumber\tagnumber=0

\immediate\newwrite\eqnfile
\newif\if@qnfile\@qnfilefalse
\def\write@qn#1{}
\def\writenew@qn#1{}
\def\w@rnwrite#1{\write@qn{#1}\message{#1}}
\def\@rrwrite#1{\write@qn{#1}\errmessage{#1}}

\def\taghead#1{\gdef\t@ghead{#1}\global\tagnumber=0}
\def\t@ghead{}

\expandafter\def\csname @qnnum-3\endcsname
  {{\t@ghead\advance\tagnumber by -3\relax\number\tagnumber}}
\expandafter\def\csname @qnnum-2\endcsname
  {{\t@ghead\advance\tagnumber by -2\relax\number\tagnumber}}
\expandafter\def\csname @qnnum-1\endcsname
  {{\t@ghead\advance\tagnumber by -1\relax\number\tagnumber}}
\expandafter\def\csname @qnnum0\endcsname
  {\t@ghead\number\tagnumber}
\expandafter\def\csname @qnnum+1\endcsname
  {{\t@ghead\advance\tagnumber by 1\relax\number\tagnumber}}
\expandafter\def\csname @qnnum+2\endcsname
  {{\t@ghead\advance\tagnumber by 2\relax\number\tagnumber}}
\expandafter\def\csname @qnnum+3\endcsname
  {{\t@ghead\advance\tagnumber by 3\relax\number\tagnumber}}

\def\equationfile{%
  \@qnfiletrue\immediate\openout\eqnfile=\jobname.eqn%
  \def\write@qn##1{\if@qnfile\immediate\write\eqnfile{##1}\fi}
  \def\writenew@qn##1{\if@qnfile\immediate\write\eqnfile
    {\noexpand\tag{##1} = (\t@ghead\number\tagnumber)}\fi}
}

\def\callall#1{\xdef#1##1{#1{\noexpand\call{##1}}}}
\def\call#1{\each@rg\callr@nge{#1}}

\def\each@rg#1#2{{\let\thecsname=#1\expandafter\first@rg#2,\end,}}
\def\first@rg#1,{\thecsname{#1}\apply@rg}
\def\apply@rg#1,{\ifx\end#1\let\next=\relax%
\else,\thecsname{#1}\let\next=\apply@rg\fi\next}

\def\callr@nge#1{\calldor@nge#1-\end-}
\def\callr@ngeat#1\end-{#1}
\def\calldor@nge#1-#2-{\ifx\end#2\@qneatspace#1 %
  \else\calll@@p{#1}{#2}\callr@ngeat\fi}
\def\calll@@p#1#2{\ifnum#1>#2{\@rrwrite{Equation range #1-#2\space is bad.}
\errhelp{If you call a series of equations by the notation M-N, then M and
N must be integers, and N must be greater than or equal to M.}}\else %
{\count0=#1\count1=#2\advance\count1 by1\relax\expandafter\@qncall\the\count0,%
  \loop\advance\count0 by1\relax%
    \ifnum\count0<\count1,\expandafter\@qncall\the\count0,%
  \repeat}\fi}

\def\@qneatspace#1#2 {\@qncall#1#2,}
\def\@qncall#1,{\ifunc@lled{#1}{\def\next{#1}\ifx\next\empty\else
  \w@rnwrite{Equation number \noexpand\(>>#1<<) has not been defined yet.}
  >>#1<<\fi}\else\csname @qnnum#1\endcsname\fi}

\let\eqnono=\eqno
\def\eqno(#1){\tag#1}
\def\tag#1$${\eqnono(\displayt@g#1 )$$}

\def\aligntag#1\endaligntag
  $${\gdef\tag##1\\{&(##1 )\cr}\eqalignno{#1\\}$$
  \gdef\tag##1$${\eqnono(\displayt@g##1 )$$}}

\def\eqalignno#1{\displ@y \tabskip\centering
  \halign to\displaywidth{\hfil$\displaystyle{##}$\tabskip\z@skip
    &$\displaystyle{{}##}$\hfil\tabskip\centering
    &\llap{$\displayt@gpar##$}\tabskip\z@skip\crcr
    #1\crcr}}

\def\displayt@gpar(#1){(\displayt@g#1 )}

\def\displayt@g#1 {\rm\ifunc@lled{#1}\global\advance\tagnumber by1
        {\def\next{#1}\ifx\next\empty\else\expandafter
        \xdef\csname @qnnum#1\endcsname{\t@ghead\number\tagnumber}\fi}%
  \writenew@qn{#1}\t@ghead\number\tagnumber\else
        {\edef\next{\t@ghead\number\tagnumber}%
        \expandafter\ifx\csname @qnnum#1\endcsname\next\else
        \w@rnwrite{Equation \noexpand\tag{#1} is a duplicate number.}\fi}%
  \csname @qnnum#1\endcsname\fi}

\def\ifunc@lled#1{\expandafter\ifx\csname @qnnum#1\endcsname\relax}

\let\@qnend=\end\gdef\end{\if@qnfile
\immediate\write16{Equation numbers written on []\jobname.EQN.}\fi\@qnend}

\catcode`@=12

\catcode`@=11
\newcount\r@fcount \r@fcount=0
\newcount\r@fcurr
\immediate\newwrite\reffile
\newif\ifr@ffile\r@ffilefalse
\def\w@rnwrite#1{\ifr@ffile\immediate\write\reffile{#1}\fi\message{#1}}

\def\writer@f#1>>{}
\def\referencefile{
  \r@ffiletrue\immediate\openout\reffile=\jobname.ref%
  \def\writer@f##1>>{\ifr@ffile\immediate\write\reffile%
    {\noexpand\refis{##1} = \csname r@fnum##1\endcsname = %
     \expandafter\expandafter\expandafter\strip@t\expandafter%
     \meaning\csname r@ftext\csname r@fnum##1\endcsname\endcsname}\fi}%
  \def\strip@t##1>>{}}

\def\citeall#1{\xdef#1##1{#1{\noexpand\cite{##1}}}}
\def\cite#1{\each@rg\citer@nge{#1}}	

\def\each@rg#1#2{{\let\thecsname=#1\expandafter\first@rg#2,\end,}}
\def\first@rg#1,{\thecsname{#1}\apply@rg}	
\def\apply@rg#1,{\ifx\end#1\let\next=\relax
\else,\thecsname{#1}\let\next=\apply@rg\fi\next}

\def\citer@nge#1{\citedor@nge#1-\end-}	
\def\citer@ngeat#1\end-{#1}
\def\citedor@nge#1-#2-{\ifx\end#2\r@featspace#1 
  \else\citel@@p{#1}{#2}\citer@ngeat\fi}	
\def\citel@@p#1#2{\ifnum#1>#2{\errmessage{Reference range #1-#2\space is bad.}%
    \errhelp{If you cite a series of references by the notation M-N, then M and
    N must be integers, and N must be greater than or equal to M.}}\else%
 {\count0=#1\count1=#2\advance\count1 by1\relax\expandafter\r@fcite\the\count0,
  \loop\advance\count0 by1\relax
    \ifnum\count0<\count1,\expandafter\r@fcite\the\count0,%
  \repeat}\fi}

\def\r@featspace#1#2 {\r@fcite#1#2,}	
\def\r@fcite#1,{\ifuncit@d{#1}
    \newr@f{#1}%
    \expandafter\gdef\csname r@ftext\number\r@fcount\endcsname%
                     {\message{Reference #1 to be supplied.}%
                      \writer@f#1>>#1 to be supplied.\par}%
 \fi%
 \csname r@fnum#1\endcsname}
\def\ifuncit@d#1{\expandafter\ifx\csname r@fnum#1\endcsname\relax}%
\def\newr@f#1{\global\advance\r@fcount by1%
    \expandafter\xdef\csname r@fnum#1\endcsname{\number\r@fcount}}

\let\r@fis=\refis			
\def\refis#1#2#3\par{\ifuncit@d{#1}
   \newr@f{#1}%
   \w@rnwrite{Reference #1=\number\r@fcount\space is not cited up to now.}\fi%
  \expandafter\gdef\csname r@ftext\csname r@fnum#1\endcsname\endcsname%
  {\writer@f#1>>#2#3\par}}

\def\ignoreuncited{
   \def\refis##1##2##3\par{\ifuncit@d{##1}%
    \else\expandafter\gdef\csname r@ftext\csname r@fnum##1\endcsname\endcsname%
     {\writer@f##1>>##2##3\par}\fi}}

\def\r@ferr{\endreferences\errmessage{I was expecting to see
\noexpand\endreferences before now;  I have inserted it here.}}
\let\r@ferences=\references
\def\references{\r@ferences\def\endmode{\r@ferr\par\endgroup}}

\let\endr@ferences=\endreferences
\def\endreferences{\r@fcurr=0
  {\loop\ifnum\r@fcurr<\r@fcount
    \advance\r@fcurr by 1\relax\expandafter\r@fis\expandafter{\number\r@fcurr}%
    \csname r@ftext\number\r@fcurr\endcsname%
  \repeat}\gdef\r@ferr{}\endr@ferences}


\let\r@fend=\endpaper\gdef\endpaper{\ifr@ffile
\immediate\write16{Cross References written on []\jobname.REF.}\fi\r@fend}

\catcode`@=12

\citeall\refto		
\citeall\ref		%
\citeall\Ref		%
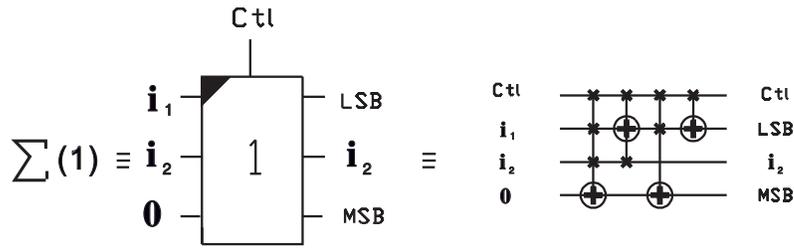
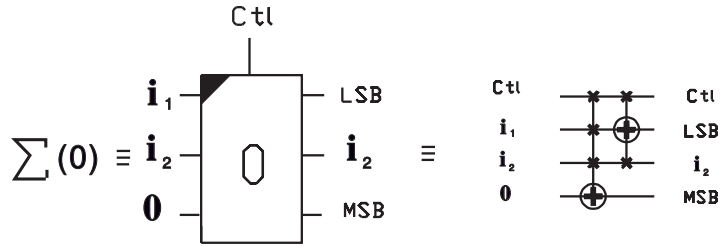
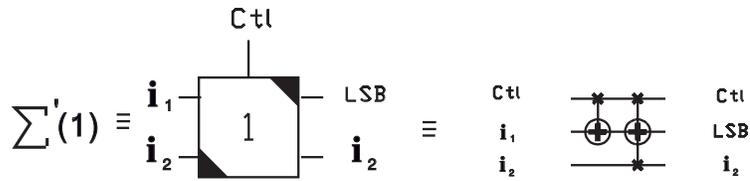
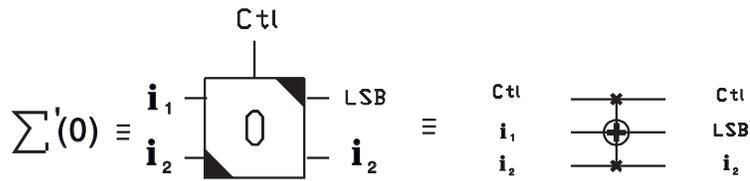
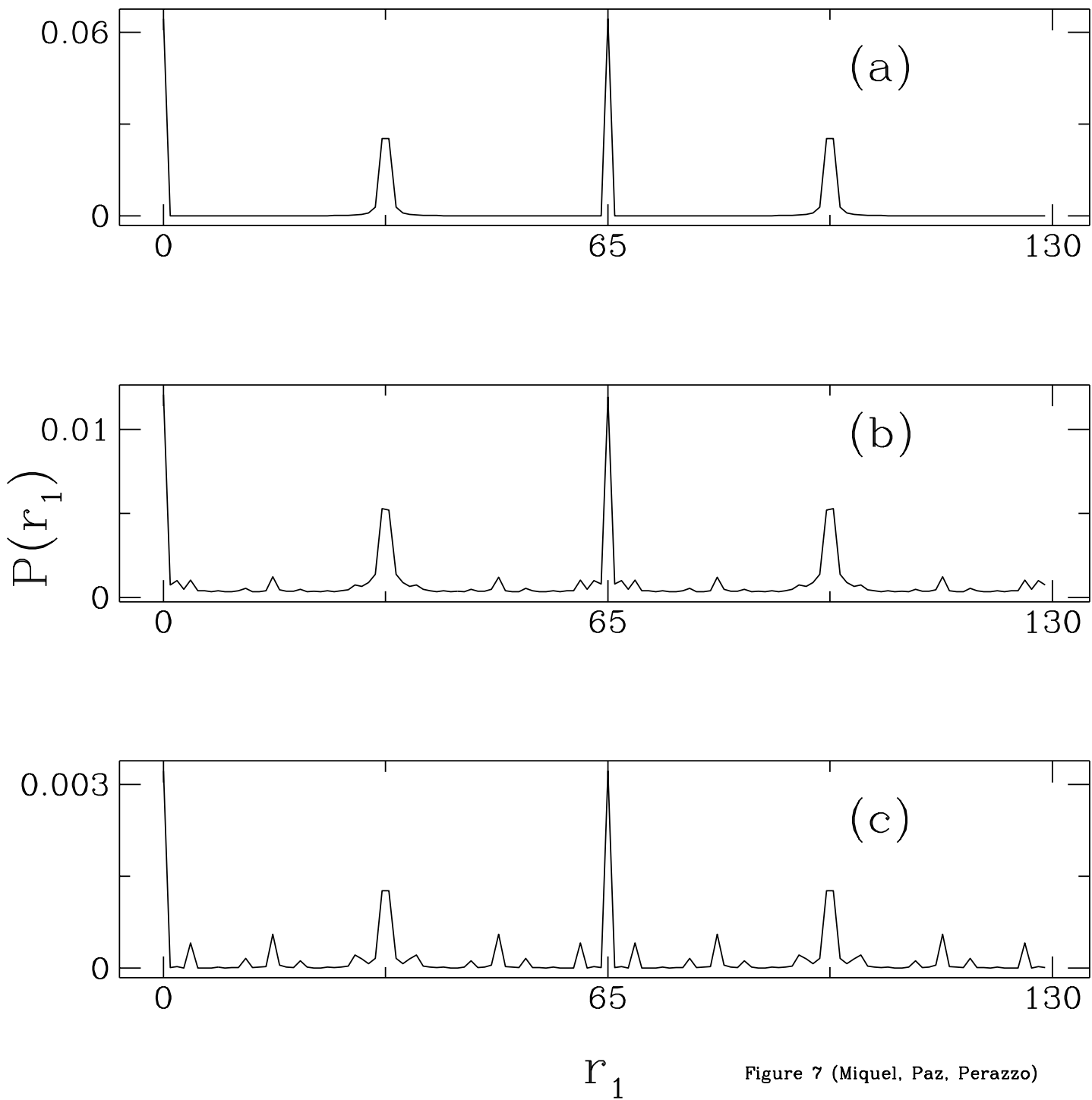
